\documentclass[twocolumn,showpacs,groupedaddress,assymb,amsmath]{revtex4}
\usepackage{graphicx}
\usepackage{amssymb}
\usepackage{amsmath}

\begin{document}

\title{Single-Photon Storing in Coupled Non-Markovian Atom-Cavity  System }
\author{H. Z. Shen, M. Qin, and X. X. Yi}
\affiliation{School of Physics and Optoelectronic Technology\\
Dalian University of Technology, Dalian 116024 China\\}
\date{\today}

\begin{abstract}
Taking the non-Markovian effect into account, we  study  how to
store a single photon of arbitrary temporal shape in a single atom
coupled to an optical cavity. Our model applies to Raman transitions
in three-level atoms with one branch of the transition controlled by
a driving pulse, and the other coupled to the cavity. For any
couplings of input field to the optical cavity and  detunings of the
atom from the driving pulse and cavity, we extend the input-output
relation from Markovian dynamics to non-Markovian one. For most
possible photon shapes, we derive an analytic expression for the
driving pulse in order to completely map the input photon into the
atom. We find that, the amplitude of the driving pulse depends only
on the detuning of the atom from the frequency of the  cavity, i.e.,
the detuning of the atom to the driving pulse has no effect on the
strength  of the driving pulse.
\end{abstract}

\maketitle
\section{Introduction}
Quantum networks composed of local nodes and quantum channels have
attracted much attention  in recent years due to a wide range of
possible applications in quantum information science
\cite{DiVincenzo1998393,Knill2001409,Cirac199959,DiVincenzo200048,Cirac199778,
Duan200367,Liu201059}, for example,  quantum communication and
distributed quantum computing. An important class of schemes for
quantum communication and computing is based on an elementary
process in which single quanta of excitation are transferred back
and forth between an atom and a photon\cite{Oxborrow200546}. This is
achieved  within the framework of cavity electrodynamics, which is
also the most promising candidate for deterministically producing
streams of single photons
\cite{Law199744,Kuhn199969,Kuhn200289,Sun200469,McKwwver2004303,Keller2004431}
of  narrowband and indistinguishable radiation
modes\cite{Legero200493}.

Dissipative dynamics of cavity-atom system has been well
investigated and deeply understood under the Markovian
approximation\cite{Carmichael1993}. This approximation  is valid
when the coupling between  system and   bath is weak such that the
perturbation theory can be applied,  meanwhile the validity of the
Markovian approximation requires that  the characteristic time of
the bath is sufficiently shorter than that of the system. However,
in practice, the coupling of  the system to bath  is not weak and
the   memory effect of the bath can not be neglected.  Typical
examples include optical fields propagating in cavity arrays or in
an optical fiber\cite{Hartmann20062,Pellizzari199779,Biswas200674},
trapped ions subjected to artificial colored
noise\cite{Turchette200062,Myatt2000403,Maniscalco200469}, and
microcavities interacting with a coupled resonator optical waveguide
or photonic
crystals\cite{Stefanou199857,Bayindir200084,Xu200062,Lin200572,Longhi200674},
to mention a few.

Previous studies of state transferring (or mapping)  between atom
and photon in cavity QED are based on Markovian approximation
\cite{Yao200595,Lukin200084,Fleischhauer2000179,Dilley201285,Hong200776}.
However, recent studies have shown that Markovian and non-Markovian
quantum processes\cite{Piilo2008100,Piilo200979,
Maniscalco200673,Bellomo200799} play an  important role in many
fields of physics, e.g.,  quantum
optics\cite{Gardiner2000,Scully1997,Walls1994} and quantum
information science\cite{Nielsen2000,Bennett2000404}. This motivates
us to explore the storing  of single photons of arbitrary temporal
shape (or a packet)  in coupled atom-cavity systems under the
non-Markovian approximation.

For this purpose,  we first extend the   input-output relation in
Ref. \cite{Dilley201285}  from Markovian system to non-Markovian
system\cite{Zhang201387}. Then we show the difference between
Markovian and non-Markovian approximations in the single photon
storing. Next we study state transfer from an input photon state to
a single-photon cavity dark state by adiabatically evolving the
system in the non-Markovian regime, the result is compared with that
given by the earlier scheme, we find that these methods are in good
agreement with each other.

The remainder of the paper is organized as follows. In Sec. {\rm
II}, we introduce a model to describe the atom-cavity system coupled
to input photons and derive the non-Markovian input-output
relations, the dynamical equations for the atom-cavity system are
also given in this section. In Sec. {\rm III}, we derive  an exact
expression  for the complex driving pulse with non-zero detunings
and non-zero populations of the excited state. In Sec. {\rm IV}, we
study the storing of the single photons taking  the non-Markovian
processes into account. In Sec. {\rm V}, we study the adiabatic
transfer via dark states between input photon and the cavity-atom
system. Discussion and conclusions are given in Sec. {\rm VI}.
\section{Equations of motion and non-Markovian input-output relations}

\begin{figure}[h]
\centering
\includegraphics[scale=0.4]{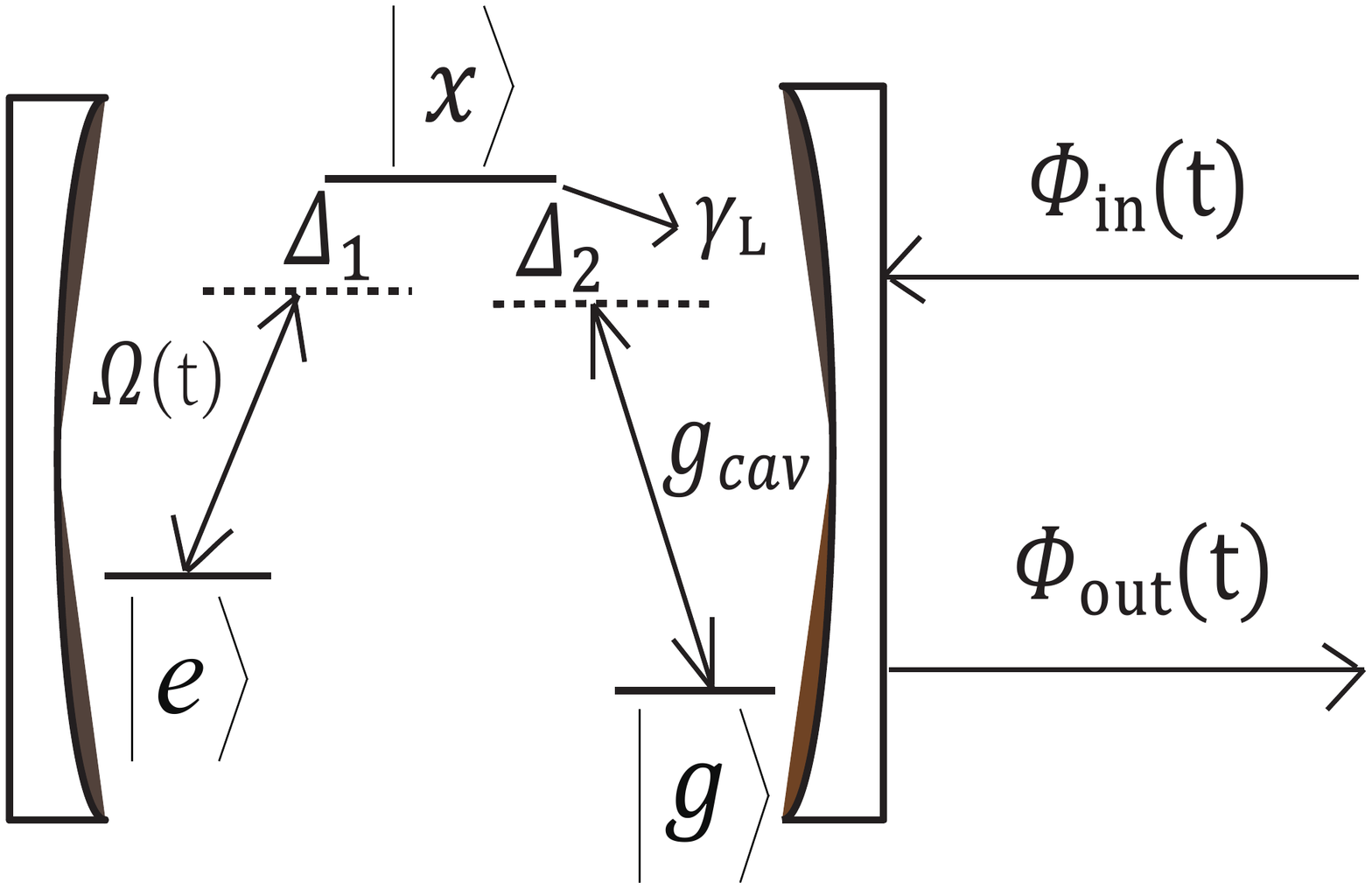}
\caption{(Color online) Schematic illustration  of our system. It
consists of a  cavity, a three-level atom, and input-output fields.
The atom is driven by both the cavity field with coupling constant
${g_{cav}}$ and a classical filed characterized by the driving pulse
$\Omega (t)$. The   classical and cavity  fields are detuning from
the  atomic resonance by ${\Delta _1}$ and ${\Delta _2}$,
respectively.} \label{figure:}
\end{figure}

We now discuss how to transfer  a  single-photon state of input
field into a single excitation of atom-cavity system. We consider an
effective one-dimensional model, which describes  a Fabry-Perot
cavity coupled to an three-level atom, as shown in
Fig.~\ref{figure:}. The input and   output fields are parallel to
the  z-axis (perpendicular to the cavity mirrors). The input field
partially transmit into the cavity through the mirror at z=0 (the
mirror at the right-hand side of the setup), the other mirror of the
cavity is assumed to be 100\% reflecting.

The input-output field is introduced as  a  continuum field modeled
by a set of oscillators  denoted by annihilation  operator $\hat
b(\omega )$, which are coupled to the cavity mode with coupling
constants $\kappa (\omega )$. The interaction between the cavity
field $\hat a$ and the continuum $\hat b(\omega )$ is described by
the following
Hamiltonian\cite{Walls1994,Gardiner198531,Fleischhauer2000179},
\begin{eqnarray}
{H_{{\mathop{\rm int}} }} = i\int_{ - \infty }^\infty   {d\omega
[\kappa (\omega )\hat a{{\hat b}^\dag }(\omega ) - H.c.} ],
\label{interaction H}
\end{eqnarray}
where $\left[ {b(\omega ),{b^\dag }(\omega ')} \right] =  \delta
(\omega  - \omega ')$ and $\left[ {a,{a^\dag }} \right] = 1$. We
consider an input field in a general single-photon state $\left|
{{\psi _{in}}(t)} \right\rangle  = \int {d\omega C_\omega
^{in}(t){{\hat b}^\dag }(\omega )\left| 0 \right\rangle } $ with
$C_\omega ^{in}(t) = C_\omega ^{in}({t_0}){e^{ - i\omega (t -
{t_0})}}$. Here, $\left| 0 \right\rangle $ denotes the vacuum state
of the continuum  $b(\omega )$. In what follows we characterize
these fields by an envelope ¡®wave function¡¯ ${\Phi _{in}}(z,t)$
defined by
\begin{eqnarray}
{\Phi _{in}}(z,t) &=& \int d\omega
\langle  {0_\omega }|\hat b(\omega ) {e^{ikz}}
\left| {{\psi _{in}}(t)} \right\rangle \nonumber\\
 &=& \int {d\omega C_\omega ^{in}(t){e^{ikz}}.}
\label{define input}
\end{eqnarray}
The normalization condition  $\int {d\omega {{\left| {C_\omega
^{in}(t)} \right|}^2}}  = 1$ of the Fourier coefficients implies the
normalization of the input wave-function according to Parseval
theorem,
\begin{eqnarray}
\int {dt{{\left| {{\Phi _{in}}(z,t)} \right|}^2}}  = 1.
\label{normalization}
\end{eqnarray}
Clearly, $\Phi _{in}(z,t)$ describes a single photon propagating
along the z-axis.

To derive an input-output relation for a general non- Markovian
quantum system, we write  the total Hamiltonian   in a rotation
frame with respect to  the center frequency ${\omega _c}$ of the
cavity field,
\begin{eqnarray}
H = {H_S} + {H_B} + {H_{{\mathop{\rm int}} }},
\label{hhh}
\end{eqnarray}
with
\begin{equation}
\begin{aligned}
{H_S} =& (\Omega (t){e^{i{\Delta _1}t}}{\sigma _{xe}}
+ {g_{cav}}{\sigma _{xg}}\hat a{e^{i{\Delta _2}t}} + H.c.) - i{\gamma _L}{\sigma _{xx}},\\
{H_B} =& \int_{ - \infty }^\infty  {d\omega {\Omega _\omega }
{{\hat b}^\dag }(\omega )\hat b(\omega )} ,
\label{hss}
\end{aligned}
\end{equation}
where ${\sigma _{\mu \nu }} = \left| \mu  \right\rangle \left\langle
\nu  \right|(\mu ,\nu  = x,e,g)$ are the atomic transition
operators, and H.c. stands for   Hermitian conjugate.  $\left| g
\right\rangle $  denotes the ground state with energy ${\omega _g} =
0$ ($\hbar  = 1$, hereafter), and $\left| e \right\rangle $ denotes
the  excited state  with energy ${\omega _e}$. ${\hat a}$ is the
annihilation operator of the cavity mode with   center frequency
${\omega _c}$. $\left| e \right\rangle $   to $\left| x
\right\rangle $ (with energy ${\omega _x}$) transition is driven by
the classical field $\Omega (t)$   with   frequency ${\nu _\Omega
}$, the transition from $\left| g \right\rangle $ to $\left| x
\right\rangle $ is driven by the cavity mode with coupling constant
${g_{cav}}$. Detuning ${\Delta _1}$ is defined as ${\Delta _1} =
{\omega _x} - {\omega _e} - {\nu _\Omega } \equiv {\omega _{xe}} -
{\nu _\Omega }$, and ${\Delta _2} = {\omega _x} - {\omega _g} -
{\omega _c} \equiv {\omega _{xg}} - {\omega _c}$. ${\gamma _L}$
denotes the atomic spontaneous emission rate and ${\Omega _\omega }
= \omega - {\omega _c}$ the  detuning of the $\omega$-mode from the
center frequency of the cavity.

Assuming   there is  only one photon initially in the input field
and the cavity-atom system is not excited, we can restrict   the
solution and discussion   of the total system (\ref{hhh})
to the subspace containing zero and a single excitation. This allows
us to expand the state vector of the total system at a later  time
$t$ as,
\begin{equation}
\begin{aligned}
\left| {\psi (t)} \right\rangle  =& G(t)\left| {g,1,0}
\right\rangle  + E(t)\left| {e,0,0} \right\rangle  + X(t)\left| {x,0,0} \right\rangle \\
 &+ \int_{ - \infty }^\infty  {d\omega {C_\omega }(t)
 {{\hat b}^\dag }(\omega )\left| {g,0,0} \right\rangle } ,
\label{total state}
\end{aligned}
\end{equation}
where   $\left| {g,1,0} \right\rangle $  denotes a state with  the
atom in the ground state $\left| g \right\rangle $, the cavity
having a single photon and   no photons in the input.   $G(t)$
denotes the probability amplitude of the total system being in
$\left| {g,1,0} \right\rangle.$  The other states have similar
notations. To calculate the probability amplitudes $G(t), E(t),
X(t),$ and ${{C_\omega }}(t)$, we substitute $\left| {\psi (t)}
\right\rangle $ into the Schr\"{o}dinger equation $i{\partial
_t}\left| {\psi (t)} \right\rangle  = H\left| {\psi (t)}
\right\rangle $. Simple calculation yields,
\begin{equation}\label{general Eq}
\begin{aligned}
\dot G =&  - i{g_{cav}}X{e^{ - i{\Delta _2}t}} -
\int_{ - \infty }^\infty  {d\omega {\kappa ^*}(\omega ){C_\omega }} ,\\
\dot E =&  - i{\Omega ^*}(t){e^{ - i{\Delta _1}t}}X,\\
\dot X =&  - i\Omega (t){e^{i{\Delta _1}t}}E -
i{g_{cav}}G{e^{i{\Delta _2}t}} - {\gamma _L}X,\\
{{\dot C}_\omega } =&  - i{\Omega _\omega }{C_\omega } + \kappa (\omega )G.
\end{aligned}
\end{equation}
Formally integrating the fourth equation of Eq. (\ref{general Eq}),
we   obtain
\begin{equation}\label{Cwt0}
\begin{aligned}
{{C_\omega }(t){\rm{ = }}{e^{ - i{\Omega _\omega }
(t - {t_0})}}{C_\omega }({t_0}) + \kappa (\omega )\int_{{t_0}}^t
{d\tau G(\tau ){e^{ - i{\Omega _\omega }(t - \tau )}}} ,}
\end{aligned}
\end{equation}
where ${C_\omega }({t_0})$ is the initial
condition of ${C_\omega }(t)$. Similarly,
\begin{equation}
\begin{aligned}
{{C_\omega }(t){\rm{ = }}{e^{ - i{\Omega _\omega }(t - {t_1})}}
{C_\omega }({t_1}) - \kappa (\omega )\int_t^{{t_1}} {d\tau G(\tau )
{e^{ - i{\Omega _\omega }(t - \tau )}}} ,}
 \label{Cwt1}
\end{aligned}
\end{equation}
where ${t_1} \ge t$. The single photon input and output  fields
${\Phi _{in}}(0,t)$ and ${\Phi _{out}}(0,t)$ (for simplicity,
hereafter we write ${\Phi _{out}}(0,t)$ as ${\Phi _{out}}(t)$,
 the same notation for ${\Phi _{in}}(t)$) are defined as the Fourier transformation
of ${C_\omega }({t_0})$ and ${C_\omega }({t_1})$ at $z=0$,
respectively.
\begin{equation}
\begin{aligned}
{\Phi _{in}}(t) =& \frac{{ - 1}}{{\sqrt {2\pi } }}
\int_{ - \infty }^\infty  {d\omega {C_\omega }
({t_0}){e^{ - i{\Omega _\omega }(t - {t_0})}}} ,\\
{\Phi _{out}}(t) =& \frac{1}{{\sqrt {2\pi } }}
\int_{ - \infty }^\infty  {d\omega {C_\omega }({t_1})
{e^{ - i{\Omega _\omega }(t - {t_1})}}} .
\label{inout}
\end{aligned}
\end{equation}
Integrating Eq.~(\ref{Cwt0}) and Eq.~(\ref{Cwt1}) and  using
Eq.~(\ref{inout}), we   obtain a  non-Markovian input-output
relation (change ${t_1} \to t$),
\begin{equation}
\begin{aligned}
{\Phi _{in}}(t) + {\Phi _{out}}(t) = \int_{{t_0}}^t {d\tau h(t - \tau )G(\tau )} ,
\end{aligned}
\end{equation}
where
\begin{equation}
\begin{aligned}
h(t) = \frac{1}{{\sqrt {2\pi } }}\int_{ - \infty }^\infty
{d\omega {e^{ - i{\Omega _\omega }t}}\kappa (\omega )} ,
\label{h}
\end{aligned}
\end{equation}
defines  the impulse response function that equals the Fourier
transform of the coupling strength ${\kappa (\omega )}$.
Substituting Eq.~(\ref{Cwt0}) into the first equation of
Eq.~(\ref{general Eq}), we obtain finally the general equations of
motion for the total system,
\begin{equation}
\begin{aligned}
\dot G =&  - i{g_{cav}}X{e^{ - i{\Delta _2}t}} + N (t) -
\int_0^t {d\tau f(t - \tau )G(\tau )} ,\\
\dot E =&  - i{\Omega ^ * }(t){e^{ - i{\Delta _1}t}}X,\\
\dot X =&  - i\Omega (t){e^{i{\Delta _1}t}}E - i{g_{cav}}
G{e^{i{\Delta _2}t}} - {\gamma _L}X,\\
{\Phi _{in}}(&t) + {\Phi _{out}}(t) = \int_{{t_0}}^t {d\tau h(t -
\tau )G(\tau )}, \label{finally Eq}
\end{aligned}
\end{equation}
where
\begin{equation}
\begin{aligned}
N (t) = \int_{ - \infty }^\infty  {d\tau {h^ * }(\tau  - t){\Phi
_{in}}(\tau )} \label{xi},
\end{aligned}
\end{equation}
is the driving field  and
\begin{equation}
\begin{aligned}
f(t - \tau ) =& \int_{ - \infty }^\infty  {d\zeta {h^*}
(\tau  - \zeta )h(t - \zeta )} \\
=& \int_{ - \infty }^\infty  {d\omega {{\left| {\kappa
(\omega )} \right|}^2}{e^{ - i{\Omega _\omega }(t - \tau )}}} \\
\equiv& \int_{ - \infty }^\infty  {d\omega J(\omega )
{e^{ - i{\Omega _\omega }(t - \tau )}}} ,
\label{f}
\end{aligned}
\end{equation}
is the memory function of the   system, and $J(\omega ) = |\kappa
(\omega ){{\rm{|}}^2}$. From the derivation, we find that $h(t)$ and
$f(t)$ plays essential roles in the photon storing. Different $h(t)$
and $f(t)$ leads to different non-Markovianity of the dynamics,
hence they affect the design of the driving pulse to store a photon
into the atom-cavity system.

\section{driving pulse  and excited state population}

In this section we present an analytical expression for the  driving
pulse to completely store  an arbitrary photon wave packet ${\Phi
_{in}}(t)$ in the atom-cavity system.  Obviously completely
impedance matching is a necessary condition for this purpose, i.e.,
\begin{equation}
\begin{aligned}
{\Phi _{out}}(t) = 0,
\label{out0}
\end{aligned}
\end{equation}
must be satisfied at any time.

The spectral response  function ${\kappa (\omega )}$ for the
Fabry-Perot (FP) cavity   can be defined by
\begin{equation}
\begin{aligned}
\kappa (\omega ) = \sqrt {\frac{\Gamma }{{2\pi }}}
\frac{W}{{W - i(\omega  - {\omega _c})}},
\label{kw}
\end{aligned}
\end{equation}
where $\Gamma $ is the cavity-input coupling strength  and $W$ is
the spectrum bandwidth of the input field. The effective spectral
density   is then \cite{Haikka201081,Breuer2002,xiong201286}
\begin{equation}
\begin{aligned}
J(\omega ) = \frac{\Gamma }{{2\pi }}\frac{{{W^2}}}{{{W^2} +
{{(\omega  - {\omega _c})}^2}}}. \label{jw}
\end{aligned}
\end{equation}
In the  wide-band limit (i.e., $W \to \infty $),  the spectral
density approximately takes $J(\omega ) \to \frac{\Gamma }{{2\pi
}}$, equivalently $\kappa (\omega ) \to \sqrt {\frac{\Gamma }{{2\pi
}}} $. This describes  the case in the Markovian limit. Then
according to Eq.~(\ref{h}) and Eq.~(\ref{f}), we have
\begin{equation}
\begin{aligned}
h(t) =& \sqrt \Gamma  \delta (t),\\
f(t) =& \Gamma \delta (t).
\label{markov}
\end{aligned}
\end{equation}
Substituting Eq. (\ref{markov}) into Eq. (\ref{finally Eq}),  we
obtain the Markovian dynamics of the total
system\cite{Dilley201285,Yao200595},
\begin{equation}
\begin{aligned}
\dot G =&  - i{g_{cav}}X{e^{ - i{\Delta _2}t}}+ \sqrt \Gamma
{\Phi _{in}}(t) - \frac{1}{2}\Gamma G(t),\\
\dot E =&  - i{\Omega ^ * }(t){e^{ - i{\Delta _1}t}}X,\\
\dot X =&  - i\Omega (t){e^{i{\Delta _1}t}}E - i{g_{cav}}
G{e^{i{\Delta _2}t}} - {\gamma _L}X,\\
{\Phi _{in}}(&t) + {\Phi _{out}}(t) = \sqrt \Gamma  G(t).
\label{markov Eq}
\end{aligned}
\end{equation}
In order to take the  non-Markovian effect into account,    we
calculate  the system-field memory  function $f(t)$ and the
spectral-response function $h(t)$
\cite{yu199960,Jack199959,Breuer2002}  by the use of  Eq.~(\ref{kw})
and Eq.~(\ref{jw}), they read,
\begin{eqnarray}
{h(t) = W\sqrt \Gamma  \Theta (t){e^{ - Wt}}}
\label{ht},
\end{eqnarray}
and
\begin{eqnarray}
{f(t) = \frac{1}{2}W\Gamma {e^{ - W\left| t \right|}}},
\label{ft}
\end{eqnarray}
where $\Theta (t)$ is the unit step function
\begin{eqnarray*}
\Theta (t) = \left\{ {\begin{array}{*{20}{c}}
{1,}&{t \ge 0},\\
{0,}&{t \le 0}.
\end{array}} \right.
\end{eqnarray*}
To store an input photon into the atom-cavity system, it is
reasonable to assume that  the total system is initially prepared in
state ${{{\hat b}^\dag }(\omega )\left| {g,0,0} \right\rangle }$,
i.e., the initial condition for the equations of motion is,
\begin{equation}
\begin{aligned}
\int {dt{{\left| {{\Phi _{in}}(t)} \right|}^2} = 1} ,
\label{phi0}
\end{aligned}
\end{equation}
\begin{equation}
\begin{aligned}
G(0) = 0,
\label{G0}
\end{aligned}
\end{equation}
\begin{equation}
\begin{aligned}
X(0) = 0,
\label{X0}
\end{aligned}
\end{equation}
\begin{equation}
\begin{aligned}
E(0) = 0.
\label{E0}
\end{aligned}
\end{equation}

Now we calculate the population of the atom in the excited state $\left| {e,0,0}
\right\rangle $,
\begin{equation}
\begin{aligned}
{\rho _{ee}}(t) =& {\rho _{offset}} - {{\tilde X}^2}(t)\\
&+ \int_0^t {dt'[2{g_{cav}}\tilde X(t')G(t') - 2{\gamma _L}{{\tilde X}^2}(t')]} .
\label{peee}
\end{aligned}
\end{equation}
Eq.~(\ref{peee}) shows that the  population of excited state ${\rho
_{ee}}(t)$ does not depend on the detunings ${\Delta _1}$ and
${\Delta _2}$. From the derivation below for the complex driving
pulse $\Omega (t)$, we  see that we should  introduce an offset term
${\rho _{offset}}$ phenomenologically to account for the imperfect
state preparation--a small initial population in the excited state
$\left| e,0,0 \right\rangle$. \textbf{We give the details of the derivations of Eq.~(\ref{peee}) in Appendix A.}

\textbf{We can now proceed to} derive the  complex driving pulse $\Omega (t)$ for completely
storing  an photon in arbitrary temporal shape  with nonzero
detunings ${{\Delta _1}}$ and ${{\Delta _2}}$,
\begin{equation}
\begin{aligned}
\Omega (t) = \alpha (t) + i\beta (t),
\label{Omega}
\end{aligned}
\end{equation}
where
\begin{equation}
\begin{aligned}
&\alpha (t) = [{\partial _t}\tilde X(t)\cos A(t) - {g_{cav}}G(t)\cos A(t)\\
& + {\gamma _L}\tilde X(t)\cos A(t) + {\Delta _2}\tilde X(t)\sin A(t)]/
\sqrt {{\rho _{ee}}(t)} ,\\
&\beta (t) = [{\Delta _2}\tilde X(t)\cos A(t) - {\partial _t}\tilde X(t)\sin A(t)\\
&+ {g_{cav}}G(t)\sin A(t) - {\gamma _L}\tilde X(t)\sin A(t)]/\sqrt {{\rho _{ee}}(t)} ,
\label{ab}
\end{aligned}
\end{equation}
with
\begin{equation}
\begin{aligned}
A(t) =  - \Delta  \cdot t + {\Delta _2}\int_0^t
{dt'\frac{{{{\tilde X}^2}(t')}}{{{\rho _{ee}}(t')}}}.
\label{At}
\end{aligned}
\end{equation}
\textbf{The details of the derivation of Eq.~(\ref{Omega}) can also be found in Appendix A.}

Modulus and argument of the complex driving pulse
 $\Omega (t)=\left| {\Omega (t)} \right|{e^{i\theta (t)}}$ is
\begin{equation}
\begin{aligned}
\left| {\Omega (t)} \right| =& \sqrt {{\alpha ^2}(t)
+ {\beta ^2}(t)}
\label{alphabeta}
\end{aligned}
\end{equation}
\begin{equation}
\begin{aligned}
=& \sqrt {\frac{{{{[{\partial _t}\tilde X(t) - {g_{cav}}G(t) +
{\gamma _L}\tilde X(t)]}^2} + {\Delta _2}^2\tilde
X{{(t)}^2}}}{{{\rho _{ee}}(t)}}}, \label{modulus}
\end{aligned}
\end{equation}

\begin{equation}
\begin{aligned}
\theta (t) = \arctan \left[ {\frac{{\beta (t)}}{{\alpha (t)}}} \right],
\label{theta}
\end{aligned}
\end{equation}
which is an analytical expression that defines the complex driving
pulse necessary to store completely the desired photon packet. This
equation tells us that  the modulus of the  driving pulse $\Omega
(t)$  depends only on the detuning ${{\Delta _2}}$, not on the
detuning ${{\Delta _1}}$.

Under the Markovian approximation \textbf{\textbf{(we denote Markovian case by introducing the subscript f)}} and defining ${X_f}(t) = -
i{e^{i{\Delta _2}t}}{\tilde X_f}(t)$ and ${E_f}(t) = {e^{ - i{\Delta
_1}t + i{\Delta _2}t}}{\tilde E_f}(t)$, we obtain from
Eq.~(\ref{markov Eq}) the following results with nonzero ${\Delta
_1}$ and ${\Delta _2}$
\begin{equation}
\begin{aligned}
{G_f}(t) =& {\Phi _{in}}(t)/\sqrt \Gamma  ,\\
{{\tilde X}_f}(t) =& [ - {{\dot G}_f}(t) + \frac{1}{2}\Gamma{G_f}(t)]/{g_{cav}},\\
{\rho _{fee}} =& {\rho _{offset}} - {{\tilde X}_f}^2\\
&+ \int_0^t {dt'[2{g_{cav}}{{\tilde X}_f}(t'){G_f}(t') - 2{\gamma _L}
{{\tilde X}_f}^2(t')]} ,\\
{\Omega _f}(t) =& {\alpha _f}(t) + i{\beta _f}(t),
\label{Markovs}
\end{aligned}
\end{equation}
where,
\begin{equation}
\begin{aligned}
{\alpha _f}(t) =& [\cos ({A_f}){\partial _t}{{\tilde X}_f}
- {g_{cav}}\cos ({A_f}){G_f}\\
&+ {\gamma _L}\cos ({A_f}){{\tilde X}_f} + {\Delta _2}
\sin ({A_f}){{\tilde X}_f}]/\sqrt {{\rho _{fee}}} ,\\
{\beta _f}(t) =& [{\Delta _2}\cos ({A_f}){{\tilde X}_f} -
\sin ({A_f}){\partial _t}{{\tilde X}_f}\\
&+ {g_{cav}}\sin ({A_f}){G_f} - {\gamma _L}\sin ({A_f}){{\tilde X}_f}]
/\sqrt {{\rho _{fee}}} ,\\
{A_f}(t) =&  - \Delta \cdot t + {\Delta _2}\int_0^t
{dt'\frac{{\tilde X_f^2(t')}}{{{\rho _{fee}}(t')}}} . \label{kiff}
\end{aligned}
\end{equation}
Within the Markovian approximation, \textbf{modulus $\left| {{\Omega _f}(t)} \right|$ and argument ${\theta _f}(t)$ of the complex driving pulse ${{\Omega _f}(t)}$ are formally the same as Eq.~(\ref{alphabeta}) and Eq.~(\ref{theta}) with the subscript $f$, i.e., $\alpha (t) \to {\alpha _f}(t)$ and $\beta (t) \to {\beta _f}(t)$.}
This driving pulse representing the coupling constant between the
atom and the driving fields is complex when the detunings are not
zero, which is not discussed in the earlier studies.

\section{single photons storing and impedance matching}
We now consider an realistic input photon packet that starts from
time ${t_{0}}$ and ends at time ${t_{e}}$. We   assume the packet
starts off smoothly, i.e.,  ${\Phi _{in}}({t_{0}}) = {\partial _t}
{\Phi _{in}}({t_{0}}) = 0$  as described in \cite{Vasilev201012}.
The second time derivative of the input $\Phi _{in}(t_{0})$ might be
nonzero at ${t_{0}}$, thus ${\rm{G(0)}} = 0$ in Eq.~(\ref{G}) but
\begin{equation}
\begin{aligned}
\dot G({t_0}) = \frac{{{{\ddot \Phi }_{in}}({t_0})}}{{W\sqrt \Gamma  }} \ne 0.
\label{Gdd}
\end{aligned}
\end{equation}
Furthermore,  from  Eq.~(\ref{X}) together with Eq.~(\ref{X0}) and
Eq.~(\ref{G0}), we find
\begin{eqnarray}
\dot G(0) =N (0),
\label{equilibrium condition}
\end{eqnarray}
this is the   so-called equilibrium condition.

By  Eq.~(\ref{xi}) and Eq.~(\ref{ht}), we can establish a relation
between $W$ and $\Gamma$ for   arbitrary input photon wave packets
${\Phi _{in}}$
\begin{eqnarray}
\Gamma  = \frac{{{{\ddot \Phi }_{in}}(0)}}{{{W^2}\int_t^\infty
{d\tau {e^{ - W(\tau  - t)}}{\Phi _{in}}(\tau )} }}. \label{gamma d}
\end{eqnarray}
We should notice that the initial conditions  from Eq.~(\ref{Omega})
now become $A(0)=0$, $\beta (0) = 0$, and  $\Omega (0) = \alpha (0)
= \frac{{{\partial _t}\tilde X(0)}}{{\sqrt {{\rho _{offest}}} }} \ne
0$. To satisfy the last initial condition, a small but nonvanishing
initial population in the state $\left| {e,0,0} \right\rangle $ is
required, in other words, perfect impedance matching with ${\rho
_{offest}} = 0$ would only be possible when the input photon packet
lasts for a very long (infinite)  time.

To exemplify the scheme and discuss  the implications of the
constraints to the initial population, we now apply the design to a
couple of typical photon shapes (or packets) that  are of general
interest. First, we consider photon wavepackets on a finite support
ranging from $0$ to $T$   symmetric in time. A particular
normalization shape (or packets) that meets the  above initial
condition is
\begin{eqnarray}
{\Phi _{in}}(t) = \frac{{8{{\sin }^2}(2t\pi /T){{\cos }^2}(t\pi /T)}}{{\sqrt {7\pi } }}.
\label{Phiin}
\end{eqnarray}
Taking $T = \pi \mu s$, we  obtain a constraint on  $W$ and $\Gamma$
in the input packet from Eq.~(\ref{gamma d})
\begin{eqnarray}
\Gamma  = \frac{{\left( {{W^2} + 4} \right)\left( {{W^2} + 16}
\right)\left( {{W^2} + 36} \right)}}{{W\left( {{W^4} + 28{W^2} + 72}
\right)\left( {1 - {e^{ - \pi {\rm{W}}}}} \right)}}. \label{gdd}
\end{eqnarray}

\begin{figure}[h]
\centering
\includegraphics[scale=0.4]{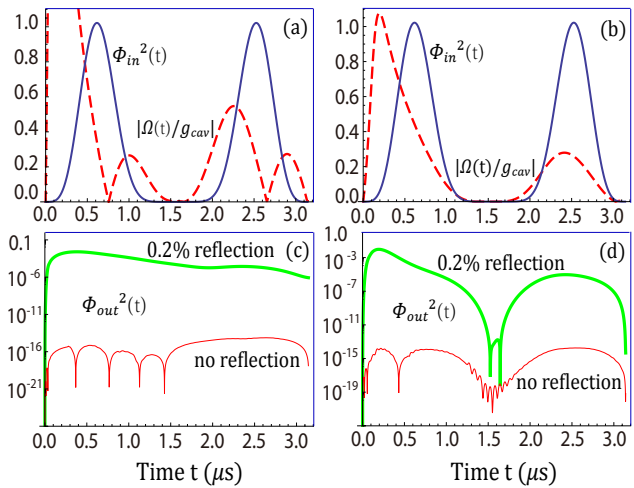}
\caption{(Color online) Input single photon packet (blue line), the
driving pulse (red-dashed line) and ${\left| {{\Phi _{out}}}
\right|^2}$ as a function of time. The parameters chosen are
${g_{cav}} = 30\pi MHz,{\gamma _L} = 6\pi MHz$. In the numerical
simulations here and hereafter, the coupling strength $\Gamma $ is
given by Eq.~(\ref{gdd}). Initially, the system is prepared
almost in $\left| {g,0,0} \right\rangle $ with a small probability
${\rho _{offset}}=0.002 $ in $\left| {e,0,0} \right\rangle $, which is plotted in (c) and (d)(thin-red lines). For
comparison, we plot ${\left| {{\Phi _{out}}} \right|^2}$  in (c) and
(d)(bold-green lines) for the case in which the system is prepared
with probability 1 in $\left| {g,0,0} \right\rangle$. Note that in
this case, we still use the driving pulse calculated with ${\rho
_{offset}}=0.002 $, so the reflection is higher.  The other
parameters chosen are ${\Delta _1}{\rm{ = }}{\Delta _2}{\rm{ = }}0$,
$W= 1.6716 MHz$ for (a) and (c), $W=17.238 MHz$ for  (b) and (d).}
\label{phiin:}
\end{figure}

Notice the unit step function in $h(t)$,   the upper and lower
limits of the integral in Eq.~(\ref{xi})  are $T$ and $t$,
respectively. For zero detunings, ${\Delta _1} = {\Delta _2} = 0$,
the   driving pulse $\Omega (t)$ (\ref{Omega}) is real. This
together with Eq.~(\ref{ab}) and Eq.~(\ref{At}) yields   $A(t) = 0$,
$\beta (t) = 0 $ and a real $ \alpha (t)$
\begin{eqnarray}
\Omega (t) = \alpha (t) = [{\partial _t}\tilde X(t) - {g_{cav}} G(t)
+ {\gamma _L}\tilde X(t)]/\sqrt {{\rho _{ee}}} , \label{realomig}
\end{eqnarray}
For an input photon packet with a duration of $T = \pi \mu s$, we
plot ${\Phi _{in}}(t)$, $\Omega (t)$, and the probability amplitude
of reflected photon, ${{\Phi _{out}}(t)}$, as a function of time in
Fig.~\ref{phiin:}. ${{\Phi _{out}}(t)}$ is obtained from  numerical
simulations of Eq. (\ref{finally Eq}) for the following two cases,
(1) the system is initially prepared in $\left| {g,0,0}
\right\rangle $, i.e.,  ${\rho _{offset}} = 0$, (2) the population
of the atom in the excited state is initially not zero (in the
figure we choose ${\rho _{offset}}=0.002$), while the cavity is
empty. We emphasize that in the numerical simulations here and
hereafter, the frequency   is re-scaled in units of $MHz$,
accordingly the time $t$ is in units of $\mu s$.  To be specific, we
choose  ${g_{cav}} = 30\pi MHz,$ and ${\gamma _L} = 6\pi MHz $ to
plot Fig.~\ref{phiin:}. This choice  of parameters was suggested in
\cite{Vasilev201012,Dilley201285}, which is within touch by current
technologies. Note that in this plot we use the same driving pulse
$\Omega (t)$, which is calculated with ${\rho _{offset}}=0.002$. We
should emphasize that the  choice of  ${\rho _{offset}}$ is
arbitrary and  limited only by practical considerations, we will
discuss this issue again later.
Fig.~\ref{phiin:} (a) and (b) show that in order to store  the input
photon completely, we have to change the driving pulse according to
the cavity-input field couplings. From Fig.~\ref{phiin:} (c), we can
learn that when the  initial state of atom matches the  conditions
used to calculate $\Omega (t)$, i.e., with ${\rho _{offset}} =
0.002$, no photon  is reflected out (it is below ${10^{ - 16}}$,
almost zero). However, if the initial state deviates from the state
used to calculate the driving pulse, say the initial state is
$\left| {g,0,0} \right\rangle $,   the photon would be  reflected
off the cavity with an  probability of $0.2\%$, which is much larger
than ${10^{ - 16}}$ and can be explained  as a  mismatch between the
initial state used to calculate the driving pulse and the realistic
initial state.

\begin{figure}[h]
\centering
\includegraphics[scale=0.4]{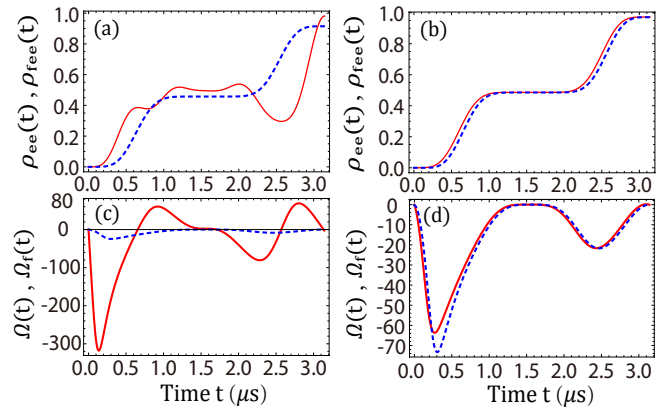}
\caption{(Color online) The populations of the atom on the  excited
state  ${\rho _{ee}}(t)$ (non-Markovian case) and   ${\rho
_{fee}}(t)$ (Markovian case), the driving pulse $\Omega (t)$
(non-Markovian case) and ${\Omega _f}(t)$ (Markovian case) versus
time $t$. The red line denotes the non-Markovian case and the
blue-dashed line denotes the Markovian case. Parameters chosen are
${\Delta _1}{\rm{ = }}{\Delta _2}{\rm{ = }}0$, ${\rho _{offset}} =
0.0075$, ${g_{cav}} = 30\pi MHz ,{\gamma _L} = 6\pi MHz. W = 0.5
MHz,$ for (a) and (c).  $W = 25 MHz$ for (b) and (d).}
\label{peeomig:}
\end{figure}

\begin{figure}[h]
\centering
\includegraphics[scale=0.4]{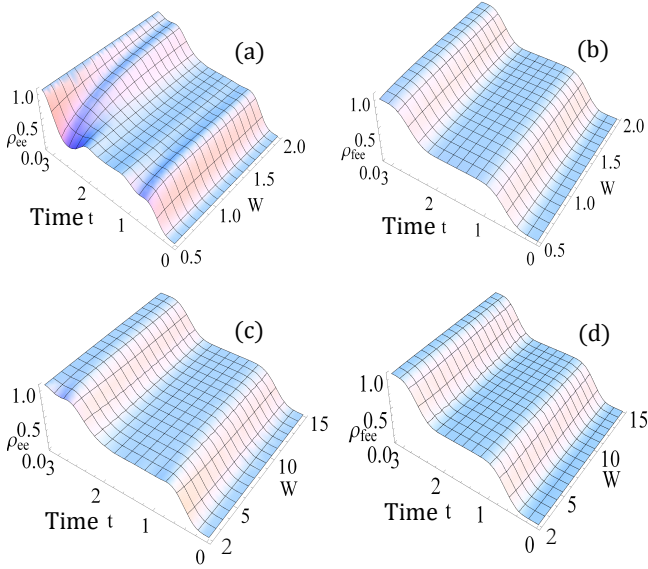}
\caption{(Color online) The time evolution of the excited  state
population in non-Markovian limit ${\rho _{ee}}(t)$ and in the
Markovian limit ${\rho _{fee}}(t)$ as a function of time $t$ and the
coupling strength $W$. Parameters chosen are ${{g_{cav}} = 30}\pi
MHz,$ ${\gamma _L} = 6\pi MHz. $} \label{3Dpee:}
\end{figure}

In order to compare the results of non-Markovian process with that
of Markovian one, we plot the time evolution of the atomic
population in the excited state $\left| e \right\rangle $ and the
real driving pulse (corresponding to zero detunings)  $\Omega (t)$
(\ref{realomig}) in Fig.~\ref{peeomig:}. We find that when the
coupling   $W$ is small (see Fig.~\ref{peeomig:} (a) and (c)), the
so-called back-flowing  phenomenon occurs for the population ${\rho
_{ee}}$. As $W$ increases,  the results given by the non-Markovian
Eq.~(\ref{peee}) are in good agreement with those given in the
Markovian limit (see Fig.~\ref{peeomig:} (b) and (d)). Besides, from
Fig.~\ref{3Dpee:} (a) and (b), we can see that the excited state
population ${\rho _{ee}}(t)$ obtained in the non-Markovian case
Eq.~(\ref{peee}) is different from that ${\rho _{fee}}(t)$ in the
Markovian case Eq.~(\ref{Markovs}) when the parameter $W$ runs from
$0.5$ to $2$, but the difference is not clear for  $W>2$, see
Fig.~\ref{3Dpee:} (c) and (d).
\begin{figure}[h]
\centering
\includegraphics[scale=0.4]{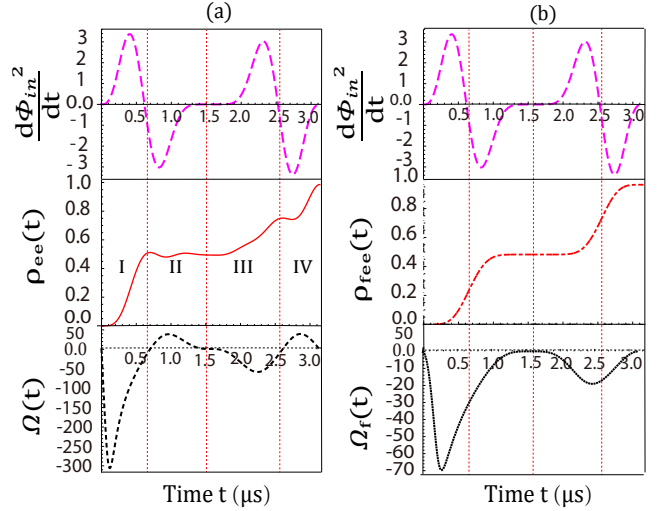}
\caption{(Color online) Comparison of   non-Markovian case to
Markovian case in terms of ${\rho}(t)$ and $\Omega (t)$. Parameters
chosen are ${\Delta _1}{\rm{ = }}{\Delta _2}{\rm{ = }}0$,
${\rm{{g_{cav}} = 30}}\pi MHz, {\gamma _L} = 6\pi MHz, W = 1 MHz,$ ${\rho _{offset}} = 0.004$.}
\label{jieshi:}
\end{figure}

To shed more light on the photon storing in  the non-Markovian
limit, we compare the non-Markovian results   with that in the
Markovian case, see Fig.~\ref{jieshi:} (a). By the input signal
${\left| {{\Phi _{in}}} \right|^2}$, we divide the dynamics  and the
time-dependence  of the driving pulse  into 4 regimes, labeled by
${\rm I}$, ${\rm II}$, ${\rm III}$ and ${\rm IV}$.  In regime ${\rm
I}$ and ${\rm III}$, the driving pulse $\Omega (t)$ is negative in
both non-Markovian and Markovian cases, while the populations of the
atom in the excited state $|e\rangle$ increase continuously in these
regimes, i.e., no population backflowing in the dynamics. In
contrary, the driving pulse in regime ${\rm II}$ and ${\rm IV}$ are
positive, and there are population  backflowing  in these regimes.

\begin{figure}[h]
\centering
\includegraphics[scale=0.4]{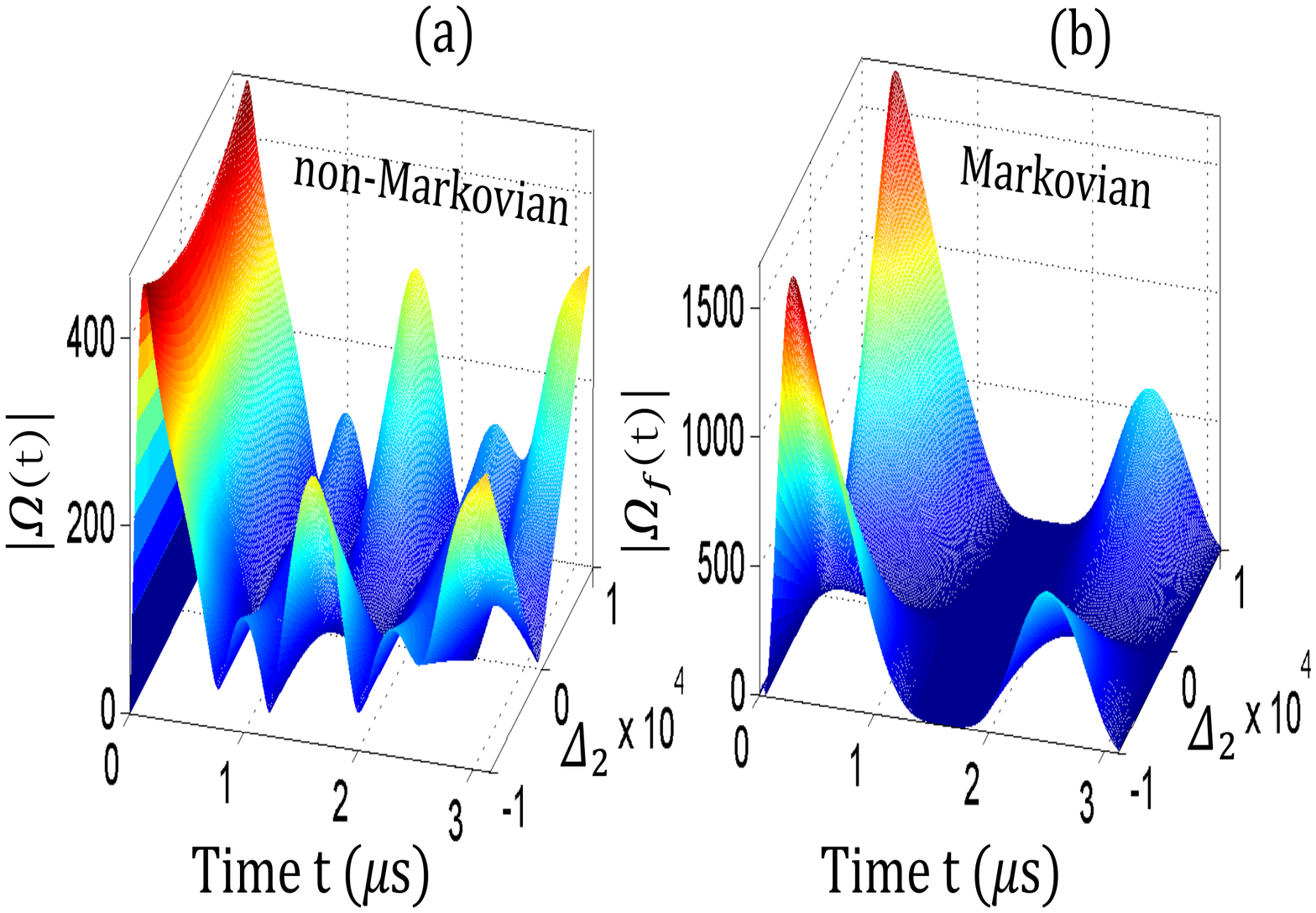}
\caption{(Color online) The modulus $\left| {\Omega (t)}  \right|$
(non-Markovian case, see Eq.~(\ref{modulus})) and $\left| {{\Omega
_f}(t)} \right|$ (Markovian case) of the
driving pulse $\Omega (t)$ vary with the detuning ${\Delta _2}$ and
 time $t$. Parameters chosen are ${\rm{{g_{cav}} = 30}}\pi MHz
,{\gamma _L} = 6\pi MHz ,W  = 0.5 MHz, {\rho _{offset}} = 0.003$.}
\label{omega:}
\end{figure}

\begin{figure}[h]
\centering
\includegraphics[scale=0.4]{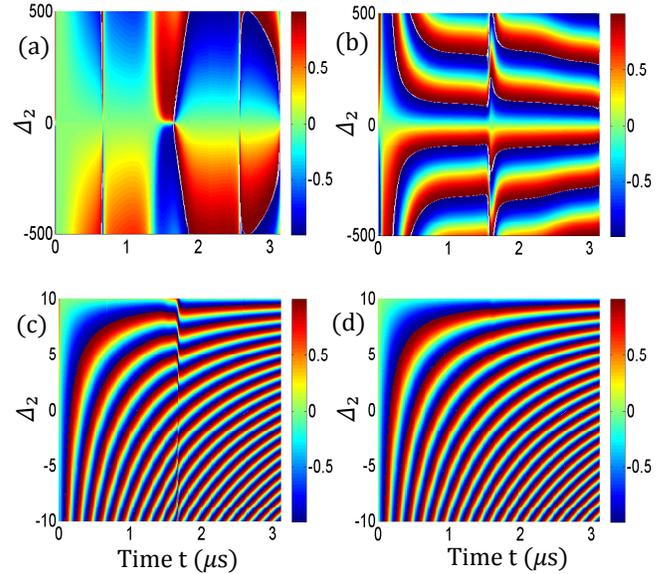}
\caption{(Color online) The argument $\sin \theta (t)$  ((a) and
(c), see Eq.~(\ref{theta})) in the non-Markovian case and $\sin
{\theta _f}(t)$ ((b) and (d)) in the
Markovian case  vary with the detuning ${\Delta _2}$ and   time $t$.
Parameters chosen are ${\Delta _1} = {\Delta _2}, {\rm{{g_{cav}} =
30}}\pi MHz ,{\gamma _L} = 6\pi MHz ,W = 0.5 MHz, {\rho _{offset}} =
0.003$, for (a) and (b), and ${\Delta _1} = 20$, for (c) and
(d).}\label{omegadelta:}
\end{figure}

Now we study the effect of detunings ${\Delta _1}$ and ${\Delta _2}$
on the driving pulse  $\Omega (t)$. Examining  Eq.~(\ref{ab}), we
find that $A(t) = {\Delta _1}t$ and $\Omega (t) = {e^{ - i{\Delta
_1}t}}({\partial _t}\tilde X(t) - {g_{cav}}G(t) + {\gamma _L}\tilde
X(t))$ when the detuning ${\Delta _2} = 0$. When ${\Delta _2} \ne
0$, the modulus $\left| {\Omega (t)} \right|$ of the driving pulse
$\Omega (t)$ does not depend on the detuning ${\Delta _1}$, while it
depends on the absolute value of ${\Delta _2}$ only (see
Fig.~\ref{omega:}). Meanwhile the argument $\theta (t)$ of the
$\Omega (t)$ depends on both  detunings ${\Delta _1}$ and
${\Delta _2}$. The argument $\theta (t)$ of the driving pulse
${\Omega (t)}$ is an odd function of  ${\Delta _2}$ (see
Fig.~\ref{omegadelta:} (a) and (b)) when  ${\Delta _1} = 0$ or
${\Delta _1} = {\Delta _2}$.
\section{Photon storing in dark states}
We now discuss the problem of transferring a single-photon state of
the input field to an atom-cavity dark state, taking  the
non-Markovian effect into account. We  show that these processes can
be achieved by adiabatically rotating the cavity dark state in a
special way. Before  proceeding,    we introduce  a dark $\left|
D(t) \right\rangle $ and its orthogonal bright states  $\left| B(t)
\right\rangle $\cite{Cohen-Tannoudji197710,Fleischhauer2000179},
\begin{equation}
\begin{aligned}
\left| {D(t)} \right\rangle  =&  - \cos
\varphi (t)\left| {g,1,0} \right\rangle  + \sin \varphi (t)\left| {e,0,0} \right\rangle,\\
\left| {B(t)} \right\rangle  =& \sin \varphi (t)\left| {g,1,0}
\right\rangle  + \cos \varphi (t)\left| {e,0,0} \right\rangle,
\label{BD}
\end{aligned}
\end{equation}
where $\tan \varphi (t) = {g_{cav}}/\Omega (t)$.

Taking the dark and bright states instead of  $\left|
{g,1,0}\right\rangle$  and $\left| {e,0,0} \right\rangle$ as the
basis, we re-expand Eq. (\ref{total state}) as,
\begin{equation}
\begin{aligned}
\left| {\psi (t)} \right\rangle  =& D(t)\left| D(t)
\right\rangle  +B(t)\left| B(t) \right\rangle  + X(t)\left| {x,0,0} \right\rangle \\
 &+ \int_{ - \infty }^\infty  {d\omega {C_\omega }(t)
 {{\hat b}^\dag }(\omega )\left| {g,0,0} \right\rangle } ,
\label{total state}
\end{aligned}
\end{equation}
The  relations between the amplitudes $D(t)$,   $B(t)$,  $G(t)$ and
$E(t)$ can be written as
\begin{equation}
\begin{aligned}
D(t) =&  - \cos \varphi (t)G(t) + \sin \varphi (t)E(t),\\
B(t) =& \sin \varphi (t)G(t) + \cos \varphi (t)E(t).
\label{dbb}
\end{aligned}
\end{equation}
The evolution equations (\ref{general Eq})   in terms of
Eq.~(\ref{dbb})  then takes (we here  consider only the ${\Delta _1}
= {\Delta _2} = 0$)
\begin{equation}
\begin{aligned}
\dot X =&  - i{\Omega _1}(t)B(t) - {\gamma _L}X,\\
\dot D =& \dot \varphi B(t) + \cos \varphi  \int
{d\omega {\kappa ^*}(\omega ){C_\omega }} ,\\
\dot B =&  - \dot \varphi D(t) - i{\Omega _1}(t)X - \sin \varphi
\int {d\omega {\kappa ^*}(\omega ){C_\omega }} ,\\
{{\dot C}_\omega } =&  - i{\Omega _\omega }{C_\omega } +
\kappa (\omega )\sin \varphi B(t) - \cos \varphi \kappa (\omega )D(t),
\label{BDequation}
\end{aligned}
\end{equation}
where ${\Omega _1}(t) = \sqrt {g_{cav}^2 + {\Omega ^2}(t)} $,  and
the terms proportional to $\dot \varphi $ describe the coupling
between the bright and dark state induced by non-adiabatic
evolutions.  We now adiabatically eliminate the excited state, this
is possible if the characteristic time ${t_1}$ of the system is
sufficiently longer than the decay  time of the excited state
(${\gamma _L}{t_1} \gg 1$). After elimination of the excited state,
we adiabatically eliminate the bright-state  and neglect terms with
$\dot \varphi $. The conditions   which validate such an elimination
will be given later. Defining $D(t) = -{d_1}$, we finally arrive
at\cite{Lukin200084,Fleischhauer2000179}
\begin{equation}
\begin{aligned}
\mathop {{d_1}}\limits^.  =&  - \cos
\varphi (t)\int {d\omega {\kappa ^*}(\omega ) {C_\omega }(t)} ,\\
{{\dot C}_\omega } =&  - i{\Omega _\omega }{C_\omega }(t)  + \cos
\varphi (t)\kappa (\omega ){d_1}(t). \label{elimination}
\end{aligned}
\end{equation}
One immediately recognizes from these equations   that the total
probability of finding the system in single photon states of the
input field and  in the cavity-dark state is conserved
\begin{eqnarray}
\frac{d}{{dt}}\left[ {{{\left| {{d_1}(t)} \right|}^2}  + \int
{d\omega {{\left| {{C_\omega }(t)} \right|}^2}} } \right] = 0.
\label{conserve}
\end{eqnarray}
Thus with adiabatic evolution,  the system can occupy only two
states, namely, the input field state and the cavity dark state.

\begin{figure}[h]
\centering
\includegraphics[scale=0.4]{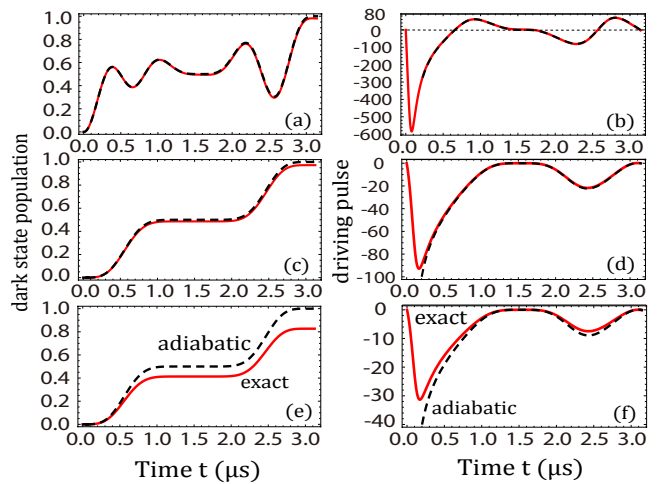}
\caption{(Color online) This plot shows the  comparison of the
adiabatic elimination approximation and the exact expression. The
red line and black-dashed line denote the exact solution (see
Eq.~(\ref{dark}) and Eq.~(\ref{realomig})) and the solution of
adiabatic elimination approximation (see Eq.~(\ref{domig})).
Parameters chosen are ${\Delta _1}{\rm{ = }}{\Delta _2}{\rm{ = }}0$,
${\rho _{offset}} = 0.00075, {\rm{{g_{cav}} = 30}}\pi MHz ,{\gamma
_L} = 6\pi MHz ,W = 0.5 MHz$ for (a) and (b), ${\rm{{g_{cav}} =
30}}\pi MHz ,{\gamma _L} = 6\pi MHz ,W = 25 MHz$ for (c) and (d) and
${{\rm{g}}_{{\rm{cav}}}}{\rm{ = 14}}\pi {\rm{MHz}} ,{\gamma _L} =
6\pi MHz ,W  = 25 MHz$ for (e) and (f).} \label{darkomig:}
\end{figure}

Formally integrating the second equation of  Eq.~(\ref{elimination})
and substituting it into the first (these steps are similar to
Eq.~(\ref{general Eq})-Eq.~(\ref{finally Eq})), we get
\begin{equation}
\begin{aligned}
\mathop {{d_1}}\limits^. (t) =& \cos \varphi (t)N(t)\\
&- \cos \varphi (t)\int_0^t {d\tau \cos
\varphi (\tau ){d_1}(\tau )f(t - \tau )} ,\\
{\Phi _{in}}(t) +& {\Phi _{out}}(t) = \int_0^t
{d\tau h(t - \tau )\cos \varphi (\tau ){d_1}(\tau )} .
\label{dphiin}
\end{aligned}
\end{equation}
We note the adiabatic evolution happens
when\cite{Lukin200084,Duan200367,Fleischhauer2000179}
\begin{eqnarray}
g_{cav}^2 \gg {\gamma _L}\Gamma .
\label{g2condition}
\end{eqnarray}
This condition is the same as that for   adiabatic storing  in the
Markovian limit, in other words, the non-Markovian and Markovian
systems share the same condition to store a photon adiabatically.
Take use of the completely impedance matching condition
Eq.~(\ref{out0}), we  obtain
\begin{eqnarray}
\cos \varphi (t){d_1}(t) = G(t).
\label{dcostehta}
\end{eqnarray}
Substituting Eq.~(\ref{dcostehta}) into the first  equation of
Eq.~(\ref{dphiin}), we  get
\begin{equation}
\begin{aligned}
{d_1}(t) =& \sqrt {2\int_0^t {M(\tau )d\tau } } ,\\
\Omega (t) =& \frac{{{g_{cav}}}}{{\tan \varphi (t)}},
\label{domig}
\end{aligned}
\end{equation}
where $M(t) = G(t){\rm{N}}(t) - G(t)\int_0^t G (\tau )f(t - \tau
)d\tau,$ $\cos \varphi (t) = \frac{{G(t)}}{{{d_1}(t)}}$. In order to
compare the analytical results  under the adiabatic evolution
Eq.~(\ref{domig})  with the exact analytical results in
Eq.~(\ref{realomig}) given by
\begin{eqnarray} {D_{dark}}(t) = G(t)\cos {\varphi _1}(t) -
E(t)\sin {\varphi _1}(t), \label{dark}
\end{eqnarray}
we plot the time evolution of the population of the dark state and
the driving pulse in Fig.~\ref{darkomig:}. Here $G(t)$ and $E(t) =
\sqrt {{\rho _{ee}}(t)} $  are the exact analytical expressions in
Eq.~(\ref{G}) and Eq.~(\ref{peee}), respectively, and ${\varphi _1}
= arc\tan [{g_{cav}}/\Omega (t)]$  is determined by
Eq.~(\ref{realomig}).
We find from the figure that the results given by the adiabatic
elimination   Eq.~(\ref{domig}) are in good agreement with those
obtained by the exact analytical expression Eq.~(\ref{realomig}) and
Eq.~(\ref{dark}) when the strong coupling conditions
(\ref{g2condition}) are satisfied (see Fig.~\ref{darkomig:} (a), (c)
and (b), (d)).   When the coupling is weak  (\ref{g2condition}) (see
Fig.~\ref{darkomig:} (e) and (f)), the curve obtained by the
adiabatic elimination approximation Eq.~(\ref{domig}) has serious
deviations from those obtained by the exact analytical expression
Eq.~(\ref{realomig}) and Eq.~(\ref{dark}). In addition, from
Fig.~\ref{darkomig:} (b), (d), and (f), we can see that the driving
pulse $\Omega (t)$ obtained by the adiabatic elimination Eq.
(\ref{domig}) shows   serious deviations  from those obtained by the
exact analytical expression Eq. (\ref{realomig}) when the time is
short (approximately $t = 0.2\mu s$),  this can be explained as an
effect of the imperfect impedance matching, in other words,
  with ${\rho _{offset}} = 0$ the perfect impedance matching
can take place only with  $\Omega(t)\rightarrow \infty$.

From Fig.~\ref{darkomig:} (a) and (b), we can  learn that the
non-Markovianity caused  backflowing  to the dark state   occurs
when the parameter $W$ is small.  The non-Markovian regime transits
to the Markovian regime when the parameter $W$ is large. Therefore
by manipulating $W$ we can control the crossover from a
non-Markovian process to a Markovian process and verse visa, this
provides us with photon storing in the atom-cavity system in both
non-Markovian and Markovian limits.


\section{Conclusion}
The storing of a single photon of arbitrary temporal shape in  a
single three-level  atom coupled to an optical cavity in
non-Markovian dynamics has been explored. To calculate the driving
pulse, we first extend the input-output relation from Markovian to
non-Markovian process, taking the off-resonant couplings between the
atom and fields into account. With the extended input-output
relation, we have presented a very simple recipe for calculating the
driving pulse with non-zero detunings ${\Delta _1}$ and ${\Delta
_2}$, and discuss the features caused by  the non-Markovian effect.
We also present a proposal to store the single photon in a dark
state of the cavity-atom system by adiabatically steering the dark
state. \textbf{In addition, due to the constraint relationship on the strength $\Gamma $ of
the coupling and the bandwidth $W$ decided by Eq. (\ref{gdd})
, we only discuss the dependence of the non-Markovian effects of the dynamics on the value of the parameter W and find that the jumping continuously from the non-Markovian regime to Markovian regime is got through manipulating width W of the band of the effective spectral density.}

\section{acknowledgments}
This work is supported by the NSF of China under Grants Nos
61078011,  10935010 and 11175032.
\appendix
\section{Calculational details of the population of the atom in the excited state and the complex driving pulse with the detunings}
\subsection{The population of the atom in the excited state}
Substituting Eq.~(\ref{ht}) and Eq.~(\ref{out0}) into the  first and
fourth equation of Eq.~(\ref{finally Eq}), we   obtain
\begin{eqnarray}
G(t) = \frac{{{{\dot \Phi }_{in}}(t) + W{\Phi _{in}}(t)}}{{W\sqrt \Gamma  }},
\label{G}
\end{eqnarray}
and
\begin{eqnarray}
\tilde X(t) = [ - \dot G(t) + N(t) - \int_0^t {d\tau f(t - \tau )G(\tau )} ]/{g_{cav}},
\label{X}
\end{eqnarray}
here, $\tilde X(t) = i{e^{ - i{\Delta _2}t}}X(t)$.  We note that the
envelope  ${\Phi _{in}}(t)$ of the  input   is a real function of
time, so both $G(t)$ and ${\tilde X}$ are real.
Defining $E(t) = {e^{ - i{\Delta _1}t + i{\Delta
_2}t}}\tilde E(t)$,   we  have from Eq.~(\ref{general Eq})
\begin{equation}
\begin{aligned}
&\Omega (t)\tilde E(t) = {\partial _t}\tilde X(t) +
i{\Delta _2}\tilde X(t) - {g_{cav}}G(t) + {\gamma _L}\tilde X(t),
\label{EE}
\end{aligned}
\end{equation}
and
\begin{eqnarray}
{\Omega ^*}(t)\tilde X(t) =  - i\Delta \tilde E(t) - {\partial _t}\tilde E(t),
\label{XX}
\end{eqnarray}
where $\Delta  = {\Delta _2} - {\Delta _1}$. It is easy to find that
${\tilde E(t)}$ and $\Omega (t)$ are complex   due to nonzero
detunings  ${\Delta _1}$ and ${\Delta _2}$, this is one of the
differences between our work and the earlier one\cite{Dilley201285}.
Taking a complex conjugation  of both sides of Eq.~(\ref{EE})
yields,
\begin{equation}
\begin{aligned}
\Omega {{(t)}^*}{{\tilde E}^*}(t) =& {\partial _t} \tilde X(t) - i{\Delta _2}\tilde X(t)\\
&- {g_{cav}}G(t) + {\gamma _L}\tilde X(t).
\label{EEE}
\end{aligned}
\end{equation}
Dividing  Eq.~(\ref{XX}) by Eq.~(\ref{EEE}), we have
\begin{equation}
\begin{aligned}
- i\Delta {\left| {E(t)} \right|^2} - {{\tilde E}^*}(t)
{\partial _t}\tilde E(t) = \tilde X(t){\partial _t}\tilde X(t)\\
- i{\Delta _2}{{\tilde X}^2}(t) - {g_{cav}}\tilde X(t)G(t) + {\gamma _L}{{\tilde X}^2}(t).
 \label{9}
\end{aligned}
\end{equation}
Taking the complex conjugation of both sides of Eq.~(\ref{XX}), we
have
\begin{equation}
\begin{aligned}
{\Omega (t)\tilde X(t) = i\Delta {{\tilde E}^*}(t) - {\partial _t}{{\tilde E}^*}(t).}
\label{10}
\end{aligned}
\end{equation}
Dividing Eq.~(\ref{EE}) by Eq.~(\ref{10}), we have
\begin{equation}
\begin{aligned}
- i\Delta {{\left| {E(t)} \right|}^2} + \tilde E(t)
{\partial _t}{{\tilde E}^*}(t) =  - \tilde X(t){\partial _t}\tilde X(t)\\
- i{\Delta _2}{{\tilde X}^2}(t) + {g_{cav}}\tilde X(t)G(t) - {\gamma _L}{{\tilde X}^2}(t).
\label{12}
\end{aligned}
\end{equation}
Using Eq.~(\ref{12}),  Eq.~(\ref{9})  and  ${\partial _t}{\rho
_{ee}}(t) = \tilde E(t){\partial _t} {\tilde E^*}(t) + {\tilde
E^*}(t){\partial _t}\tilde E(t)$, we get a differential equation of
${\rho _{ee}}(t)$
\begin{equation}
\begin{aligned}
{{\dot \rho }_{ee}}(t) =&  - 2\tilde X(t){\partial _t}
\tilde X(t) + 2{g_{cav}}\tilde X(t)G(t)\\
&- 2{\gamma _L}{{\tilde X}^2}(t).
\label{pee}
\end{aligned}
\end{equation}
\textbf{Therefore, Eq.~(\ref{peee}) is obtained by formally integrating Eq. (\ref{pee}).}
\subsection{The complex driving pulse with the detunings}
Multiplying  both
sides of Eq.~(\ref{12}) by $-i$ and taking   the complex conjugation
of the result, we obtain
\begin{equation}
\begin{aligned}
- \Delta {{\left| {\tilde E(t)} \right|}^2} + i{{\tilde E}^*}(t)
{\partial _t}\tilde E(t) =  - i\tilde X(t){\partial _t}\tilde X(t)\\
- {\Delta _2}{{\tilde X}^2}(t) + i{g_{cav}}\tilde X(t)G(t) -
i{\gamma _L}{{\tilde X}^2}(t).
\label{14}
\end{aligned}
\end{equation}
Considering
\begin{equation}
\begin{aligned}
{{\tilde E}^*}(t) =& \frac{{{\rho _{ee}}(t)}}{{\tilde E(t)}},\\
{\rho _{ee}}(t) =& {\left| {\tilde E(t)} \right|^2},
\label{144}
\end{aligned}
\end{equation}
substituting Eq.~(\ref{144}) into Eq.~(\ref{14}) and formally
integrating the obtained result from $0$ to $t$, we arrive at
\begin{equation}
\begin{aligned}
&\tilde E(t) = \tilde E(0)\exp \int_0^t {dt'\{ [\Delta {\rho _{ee}}(t')
- i\tilde X(t'){\partial _{t'}}\tilde X(t') - } \\
&{\Delta _2}{{\tilde X}^2}(t') + i{g_{cav}}\tilde X(t')G(t')
- i{\gamma _L}{{\tilde X}^2}(t')]/(i{\rho _{ee}}(t')\},
\label{finallyE}
\end{aligned}
\end{equation}
where  $\tilde E(0) = \sqrt {{\rho _{offset}}}$ representing the
initial offset, i.e., the probability amplitude of finding the
system in the  excited state. \textbf{Finally, we can obtain Eq.~(\ref{Omega}) by substituting Eq.~(\ref{finallyE}) into Eq.~(\ref{EE}) and separating the real and imaginary part of the complex driving pulse ${\Omega (t)}$.}

\end{document}